\documentclass[12pt]{article}
\usepackage{color}
\usepackage{geometry}
\definecolor{navyblue}{rgb}{0.0, 0.0, 0.5}
\usepackage{hyperref}
\hypersetup{
colorlinks=true,
urlcolor=blue,
linkcolor=navyblue
}
\begin{document}
\title{\textbf{The Decrits Consensus Algorithm:} \Large{Decentralized Agreement without Proof of Work}}
\author{Ryan Pfeiffer \href{mailto:ryan26ix@yahoo.com}{ryan26ix@yahoo.com} 
\normalsize{\hyperlink{pgp}{PGP Key}}\\
\url{www.decrits.org}}
\date{November 4, 2014}
\maketitle
\abstract{Decrits is a cryptocurrency\footnote{\textit{Cryptocurrency:} A digital construction of classical money that is protected from duplication and fraud by cryptography.} in development that makes use of a novel consensus algorithm that does not require proof-of-work\footnote{\textit{Proof-of-Work:} An algorithm which provides a computationally infeasible problem to solve, modified by a difficulty factor to allow partial solutions. These solutions are easy to verify and prove the performance of computational work.}. This paper describes how the Decrits Consensus Algorithm \textit{(DCA)} is as trustless as a proof-of-work algorithm while offering superior transaction security at virtually no cost.}
\newpage
\tableofcontents
\newpage
\section{Introduction}
Satoshi Nakamoto's landmark paper, \textit{Bitcoin: A Peer-to-Peer Electronic Cash System}\cite{bitcoin}, showed that a decentralized, trust-free, digital cash protocol can solve the \textbf{double-spending problem}\footnote{\textit{Double-Spending Problem:} The concept of a decentralized, digital currency raises the question: ``In the absence of a central authority, who or what determines the correct sequence of events in order to prevent spending the same money twice?''} by using a concept called \textbf{proof-of-work} \textit{(POW)} as long as greater than 50\% of the work is performed by honest members of the network. While at least one paper has described some weaknesses of this concept\footnote{Ittay Eyal and Emin G\"un Sirer, \textit{Majority is not Enough:
Bitcoin Mining is Vulnerable}, \url{http://arxiv.org/pdf/1311.0243v2.pdf}, 2013}, it remains a stalwart system. However, the capital costs of maintaining this system are rather exorbitant--currently estimated at nearly \$800 million USD annually\cite{coindesk} while the Bitcoin total network capitalization is around \$5 billion USD.

Recognizing this conundrum, the first proposal for a concept called \textbf{proof-of-stake} \textit{(POS)} eliminated the need for POW by relying on users of the currency to vouch for a given order of events, weighted by the amount of currency they possess\cite{pos}. Unaware of the proposal, the author of this paper proposed a vaguely similar but naive implementation several months later as part of an ongoing cryptocurrency design paper. This eventually evolved into the DCA.

Since then, several cryptocurrency implementations that do not rely on POW have made it to market: \textbf{Peercoin} which uses the Bitcoin code base and adds POS functionality very similar to the first proposal; \textbf{Ripple} which uses a hidden set of public nodes trusted not to collude; and \textbf{Nxt} which again uses a similar system to the first proposal but provides an entirely new code base. The security of each of these protocols has received passionate dispute at one time or another, with little academic research published as of yet.

The DCA is thematically similar to the prior proposals and implementations, but it aims to solve the perceived problems therein such as reduced decentralization and/or ``nothing at stake'' where no significant punishment exists for stakeholders who (seemingly) defy the rules of the network.

This paper is an outline of only the DCA and not the Decrits protocol as a whole. Concepts that are not specifically necessary to describe the DCA's operation will be discussed only minimally. Among others, a paper that describes a significant reinforcement to the strength of the DCA in concert with the Decrits network protocol will be released in the future.

\section{Basic Concepts}
\subsection{Records}
A \textbf{Record} is the basic building block of network consensus provided by the DCA. It is roughly equivalent to the Bitcoin \textbf{block} and contains a set of transactions along with other network data. In lieu of a POW solution, a Record is created and digitally signed\footnote{\textit{Digital Signatures:} \url{http://en.wikipedia.org/wiki/Digital_signature}} by a network peer selected by the DCA in a particular order--one for every 10 seconds of real time.

Each Record acknowledges as many Records as possible prior to itself, chaining them together. A new Record needs only to confirm Records that have not already been confirmed in the chain.
\subsection{Voices}
A \textbf{Voice} is a person (or group of people) who has chosen to participate in the network consensus by risking a predefined amount of Decrits currency in return for a portion of transaction fees collected by the network. To receive these fees and to avoid penalties, a Voice must create Records at the predetermined time decided by the DCA. A Voice may also be referred to as a network object, similar to a peer or a node.

A Voice is not an ephemeral concept; the currency risked cannot be used for any other purpose for a period of approximately one year. There is no limit to the amount of Voices that may exist; it is governed only by the total amount of currency people wish to risk to protect the network.

Becoming a Voice is considered a risk for two primary reasons: the opportunity cost of losing the liquidity of the currency during the period in which it is locked, and the possibility of the computer performing the Voice's duties being compromised by malware that maliciously creates invalid Records. A Voice who creates two Records for the same window of time or includes an invalid transaction\footnote{\textit{Invalid Transactions:} Transactions with an incorrect signature or for an amount greater than the account possesses.} is considered in violation of the network protocol and will have its risked Decrits destroyed.
\subsection{Consensus Blocks and Cycles}
A \textbf{Consensus Block} \textit{(CB)} is a snapshot of all the Decrits protocol data at a given point in time that is taken approximately every 10 days. This is a reference point for which nodes new to the network or ones that have not connected in some time may refer to as a starting point for network data. Having an up-to-date picture of the network data is as simple as downloading the snapshot and all subsequent Records since the snapshot. As part of a Record, a Voice will include a hash\footnote{\textit{Hash:} \url{http://en.wikipedia.org/wiki/Cryptographic_hash_function}} of the prior CB to confirm its approval.

A \textbf{Consensus Cycle} \textit{(CC)} is the period over which data is aggregated before determining a new CB. CCs are not explicitly separated by CB snapshots, but lag several days behind to allow for Voices who may have missed creating a Record to still confirm the prior CB. An additional period of about a day is provided to allow for a new order of Records (see \hyperlink{oor}{section 5.1}) to propagate the network, giving Voices time to prepare for the next cycle.
\section{The Voice Ledger}
\subsection{Description}
The defining concept of the DCA is known as the \textbf{Voice Ledger} \textit{(VL)}. The VL is a record of all Voices who have ever existed at any point in the network's history. Whenever someone opts to become a Voice, the transaction that executes this action will be added to the VL. Whenever someone no longer wishes to remain a Voice, the transaction executing this action will also be added to the VL.

Part of the reason for requiring a lengthy minimum period of time to be a Voice is to keep the VL compact and easy to transmit--there is less turnover among Voices. Its fundamental purpose acts similarly to the combined difficulty of the Bitcoin block-chain: it protects the history of the network. However, while forging Bitcoin's chain only requires creating a new history of work\footnote{\textit{Forging Bitcoin's block-chain:} Make no mistake, this is an incredibly difficult task. However, it can be much easier against less well-protected networks that rely on POW.}, forging the VL would require catastrophically breaking the underlying digital signature algorithm\footnote{\textit{Breaking the DSA:} Most modern signature schemes' security rely on problems that are currently believed to be unsolvable, but may be solved in the future.}.

Like Bitcoin's genesis block, the Decrits software will contain a genesis CB that contains a list of public keys in the VL representing the initial Voices of the network. In lieu of anonymous POW, these Voices are tasked with creating and protecting the network history.

\subsection{Defending the Voice Ledger}
While it is intuitively possible to create a forged network history by controlling some of the Voices' keys, the properties of the VL make it so that these attacks are rendered entirely ineffective.

The primary attack vector on the VL is to create an alternate network history in which the Voices under the attacker's control become the only Voices securing the network. This attack is fairly simple to execute and could even rewrite the entire history of the network if the attackers control any of the Voices in the genesis CB.

However, a key concept of the VL is that when a Voice chooses to end his tenure, he will sign a transaction stating so, and this transaction is added to the VL. This transaction is the key that unlocks the Decrits that were originally risked to protect the network. Without it, the Voice does not retrieve his money. Any rewriting of history cannot possibly include these transactions for Voices which the attacker does not control.

\textbf{It then becomes \underline{obvious}\footnote{\textit{Obvious:} No human or other external input required to determine the correct network.} to any peer new to the network which fork\footnote{\textit{Fork:} One of two or more competing network histories vying for dominance.} is honest: the one that contains these signatures.} The signatures cannot be carried over from one fork to the other because the exit transaction includes a recent CB hash which will be invalid on the dishonest fork. If the Voices in question are still working for the network, recent signatures will be available via the Records they have created.

\section{Confirming \& Securing Transactions}
\subsection{Recapping Bitcoin's Security}
A significant drawback of the Bitcoin network is the amount of time it takes to confirm a transaction. Requiring anywhere from a few minutes to occasionally longer than an hour, even when a transaction is \textit{confirmed}, it is not necessarily \textit{secure}. From the \url{bitcoin.org} website: ``For larger amounts like 1000 US\$, it makes sense to wait for 6 confirmations or more. Each confirmation \textit{exponentially} decreases the risk of a reversed transaction.''\footnote{\url{https://bitcoin.org/en/you-need-to-know}} [original emphasis]

The website, by virtue of the protocol as a whole, makes no guarantee of transaction security. The statement of ``Each confirmation exponentially decreases the risk of a reversed transaction'' is also technically incorrect. It assumes that an attacker with significant resources can only control some significant fraction of the Bitcoin hashing power that is less than 50\%, but never equal to or greater than 50\%.

While this attack seems unlikely, and indeed is a weakness inherent to POW that is conceded by Bitcoin, it \textit{is} possible and must be considered. The only defense is to wait until enough resources have been consumed by the network to exceed the value of the transaction\footnote{From a game-theoretic point of view, it only makes sense for the attacker to perform this attack if they can manage a profit.}--and even then this security is weakened by the probability of the attacker executing multiple attacks at once, multiplying the profitability of the attack.

In the future when Bitcoin mining is no longer as heavily subsidized as it is today, the problem is compounded by the fact that the network's security is relative to the amount of transaction fees assessed. If transaction fees are low, the comparative network security must be low. If transaction fees are high, it is likely that fewer transactions will be performed \textit{on-chain}\footnote{\textit{On-Chain Transactions:} Transactions that are sent directly over the network rather than through a third-party service such as a wallet website.} which would also contribute to lower network security and an easier opportunity to attack. It may very well be the case that older, less efficient pieces of silicon known as ASICs that can only perform Bitcoin POW may find new life and profitability in attacking the Bitcoin network.

\subsection{The Decrits Security Model}
In contrast to the somewhat hazy properties of the Bitcoin model, the Decrits transaction security model is simple:\\\textbf{Any transaction appearing in a Record is confirmed, and any given Record that is confirmed by at least 10 subsequent Records is secured.}

\textit{Confirmed} means that a Voice has ``wagered'' all of the Decrits he has on deposit that the transaction is valid based on his view of the network, and \textit{Secured} means that reversing the transaction could only be achieved by an unresolvable network fork\footnote{\textit{Unresolvable Network Fork:} A fork in the network that will likely remain permanent. Assuming both forks continue to exist (one does not die out), each fork will eventually destroy the other's Voices' money, and the Decrits user will have to choose which one to use. See \hyperlink{resfork}{section 5.3}.}. This means that at the very least one Voice has forked the network in a way that nodes who were not watching the network at the time require human input (in some cases) to make a decision.

The figure of 10 Records for security is a variable intended to give the network enough time to propagate data. Should a newer Record fail to confirm an older Record that has already been confirmed by 10 Records in between, \textbf{and} the ``unconfirming'' Record contains a double-spent transaction, the newer Record will create an unresolvable network fork. 

However, an honest Voice will not blindly create a Record it knows could cause a fork when its network connection might be at fault--the likely scenario is that it is under attack.  Instead, when the situation does arise, it will create a \textbf{Safe Record} that acknowledges the prior Records it has available, but does not confirm any transactions. Safe Records pose a minor but necessary nuisance by further delaying transaction confirmations.

In a network where Voices are present and timely, transactions will take between 5-15 seconds to confirm and 105-115 seconds to secure. 
\\\\
The following subsections describe some of the possible malevolent scenarios that may arise in the security model.
\subsubsection{Scenario: Double Spending Insecure Transactions}
The simplest way to avoid any possibility of having a transaction double spent against a Decrits network user is to wait until the Record containing the transaction in question has been confirmed by at least 10 subsequent Records. However, many times this is not practical, such as in the case of point-of-sale transactions where confirmation speed on the order of seconds is important.

In this case the scenario arises where a Voice could purchase a product in person in during the time window of his Record and execute a double spend on an insecure transaction. The Voice does not create his Record on time and waits for the transaction to be confirmed by a Record \textit{after} his Record's window of time. He then creates his Record with a different transaction that sends the currency to a different account, and submits it to the network.

In theory, because his Record is earlier in the order of Records than the Record that confirmed the real transaction, his Record should take precedence and the latter transaction should be considered invalid. In practice, network nodes that have seen both Records in the order they actually appeared will refuse to transmit the Record that is obviously attempting a double spend. Without controlling a huge majority of the network's nodes, convincing the network that the double spent transaction came first will be a nearly impossible task.

However, it is not a \textit{completely} impossible task. If the Voice in question also controls a large portion of the subsequent Records, he could order the transactions regardless of what the network thinks. The network has to accept this. This means that if a merchant suspects anything out of the ordinary, \textbf{and} a recently past Record is missing, it would be prudent to wait until the transaction is secured.
\subsubsection{Scenario: Isolating a Voice}
A malevolent entity may be able to isolate a Voice from the rest of the Decrits network. Provided that this is possible, by creating a public set of Records and a secret set of Records fed only to the targeted Voice, the Voice could be convinced to unknowingly unconfirm another Record or confirm a double spent transaction.

This would require an unlikely confluence of malevolent Voices to be aligned together in the order of Records, positioned just prior to the Voice being targeted. Mitigating factors to this attack include the destruction of all malevolent Voices' Decrits for creating multiple Records for the same window, the difficulty and unlikelihood of organizing malevolent Voices to be in the proper position to perform this attack, and further protection provided by the network protocol to be described in a future paper.

The probability of executing this attack is very low, the attacker will lose far more money than the victim (unless they can string together a significant number of victims, significantly increasing the attack's difficulty), and the possibility exists that the victim has an outside source of information thereby easily catching the attacker in the act--the attacker will lose his money but the victim will not.
\hypertarget{itn}{}\subsubsection{Scenario: Isolating a Transacting Node}
Isolating a transacting node may be \textit{somewhat} easier to execute, primarily for transactions done in person. If, for example, the transacting node is a cell phone in a bunker underground where the only Internet access is provided by wifi under the control of the men in suits across the table, they may be able to control the network view.

Now if the men in suits are able to predict--days in advance--within a few minutes of when the transaction will occur, and by luck or by rigging the network they are able to control the Voices for that period of time, and they are sure the victim will accept the transaction as being complete before running out of consecutive Voices to create Records, then and only then could a double spend be executed.

In certain, \textit{less-than-ideal} situations, it may be prudent to wait additional time to be sure that the Record chain has not been hijacked. The attackers will still lose Decrits for each Record that the victim waits as they must create both a secret and a public Record, and if the victim becomes aware at any point before parting with merchandise that the chain is a fraud, he can end the transaction and \textit{still} prove the duplicate Records to the network. This assumes he is able to leave the bunker, however.
\subsubsection{Scenario: Partitioning the Decrits Network}
Partitioning the Decrits network may be possible, most likely by partitioning the Internet itself such as the case in Egypt in 2011. This may cause an unresolvable network fork if the portion of the network partitioned is large enough or the partition remains in place for a long enough time.

Without a free and open internet, all cryptocurrencies are at the mercy of those who control the flow of information. Technologies such as wireless mesh networking\footnote{\url{https://en.wikipedia.org/wiki/Wireless_mesh_network}} and even satellite backup transmission are of paramount importance to protect the network from this attack. Abandoning complete Internet neutrality\footnote{\url{https://en.wikipedia.org/wiki/Net_neutrality}} is also a threat. None of these threats can be conquered within the scope of the network; it is a social and political issue.

A novel property of the Decrits network is that all users will be able to relatively quickly detect a network partition by noting the missing Records. If many more Records are missing than normal, the client should warn the user of the partition possibility.
\subsubsection{Scenario: Voices Colluding to Fork the Network}
A group of malevolent Voices could collude in secret or in public to unresolvably fork the network.

If done in secret, all nodes of the network that are connected during the time in which the attack takes place will know with certainty which fork is honest simply by observing and noting the great time disparity between receiving the forks.

If done in public, all nodes of the network that are connected during the attack will know with certainty which fork is honest because the earliest, on-time Record in order will take precedence.

Nodes that were not monitoring the network during this period of time will not know which fork is honest. This is a concession that is made by the DCA. There is a keen difference here between Decrits' concession and Bitcoin's concession (regarding 50\% attacks): \textbf{the Decrits user is not affected by the result}. If they were not watching, they were not conducting commerce with Decrits and could not be fooled into believing in a double spent transaction. Any commerce that the user is interested in after this point is likely to be completely unaffected and secured on both forks.

On the other hand, a Voice is always affected as it must do its part to help the honest network and confirm the honest fork. It may issue a Safe Record to divert the decision to a later time, but the decision must be made, and it must be made with common-sense, human input. In a ubiquitous network that Decrits would hope to be, if \textit{you} don't know what happened, then your friend or your neighbor does, or the mom and pop shop down the street does. \textbf{Calling for common-sense, human input is a small concession to make when the benefit is the impossibility of double spending and true security of transactions.}
\section{Additional Information}
\hypertarget{oor}{}\subsection{Records and Consensus Cycles}
Each CC, a new order of Records is created by using a random number (see \hyperlink{grn}{section 5.4}) as a seed. Via this mechanism, the Voices have no control over when they will be tasked to create Records. Should malevolent Voices be able to control the order of Records, they would be able to organize Records that they control together, leading to the possibility of attacks such as the one covered in \hyperlink{itn}{section 4.2.3}.

The actual Decrits CC length is 876,600 seconds (about 10.15 days), and the time between Records is 10 seconds, so the number of Record ``slices'' is 87,660. If there are fewer Voices than this number, some or all Voices will be tasked with creating multiple Records. If there are more, more than one Voice will create a Record for some or all slices. When there are two or more Voices assigned to one slice of network time, a modulo algorithm will determine which Voice records which transactions so that they do not duplicate each other's work.
\subsection{Network Time}
Transactions on the Decrits network are timestamped. Each Record slice covers a specific window of time. Without a highly coordinated attack, it will be nearly impossible to confuse a node's sense of network time. If a node is ever confused, the client will require user input on timestamping a transaction. In the case of a Voice, the client will most likely issue a Safe Record.
\hypertarget{resfork}{}\subsection{Resolving Unresolvable Forks}
An unresolvable fork does not imply the network can never be whole again, it means that a decision must be made regarding network consensus.

The most extreme example imaginable is presented: All Voices except for one lone Voice are corrupt. All users of the Decrits network watch the network at all times. The corrupt Voices double spend a transaction. The honest Voice carries the network on by itself, destroying all of the corrupt Voices' money by refusing to confirm their Records. The entire network, save the corrupt Voices, agrees without any human input required. If all nodes watch the network at all times, the network is secure from attack as long as one Voice is honest. Transaction confirmations will be temporarily and significantly delayed.

Of course both sides of the scenario are unrealistic. Nodes not watching the network will be confused about which network is correct and \textbf{cannot} make the assumption that the network with more confirmations is correct. The human in control of that node must ask others. This could be sanctioned in software (e.g. `I trust Google's version'), or it must be resolved by seeking out the accounts of others. Resolving this issue is not a matter of \textbf{trust}, it is a matter of \textbf{common sense}. A few players at the table cannot collude when a large faction of the world has seen their hole cards.
\\\\
On the other side of the coin, not all unresolvable forks are intentional and malicious. Partitioning the Internet will likely cause one. Because of this, each fork does not immediately destroy the decrits of the Voices on the other fork. Instead, the Voices in question are ``jailed''--they may not retrieve their money or create Records while in jail. In the case of a non-malicious fork, a consensus among Decrits users must be made to select which fork is the primary fork. This will probably be the fork that has a larger portion of the Voices. If only a single country is partitioned away, the rest of the world should rightly expect that their fork is the primary.

Under this scenario, the primary fork must \textbf{vote} to pardon the jailed Voices. The voting system will be described in a future paper.
\hypertarget{grn}{}\subsection{Generating a Random Number}
During each CC, each Voice will include in a Record an encrypted string and a decryption key. The decryption key will decrypt the string from the Voice's Record in the prior CC, and this string will be hashed together with all of the other Voices' strings to generate a random set of data. This random hash will be used to create a new, random order of Records in the next CC. It will also be used for other purposes to be described in future papers.

To affect the output of this generator, a group of colluding Voices could withhold their Records until the very end of the CC. If the group controls the \textbf{very last Record} of a CC, they could see all outputs of the generator by providing or continuing to withhold the Records. For this reason, a Voice that does not supply a Record at all during a CC will have its money destroyed. This turns a relatively free attack into a rather costly one.
\subsection{Attacking Lite Clients}
A \textbf{lite client}, known as a \textbf{Simplified Payment Verification} \textit{(SPV)} node in Bitcoin parlance, chooses to receive only Record headers in lieu of all network data. When receiving a transaction, the sender is expected to provide proof that the transaction exists in a Record for the lite client to accept the payment.

In Bitcoin, it is quite difficult but possible to create a fake history that is fed only to an isolated SPV node in order to double spend. In Decrits, this attack is as difficult to perform as the one described in \hyperlink{itn}{section 4.2.3}.
\subsection{Future Papers}
The DCA is a critically important concept of the Decrits network, but it is only a part of the whole. This paper has not described at all the operation of the actual network protocol, the monetary system, or the voting system. The author believes that each of these categories has received an equal amount of thought and attention, and each deserves its own paper. These will be released during the continued development of Decrits.
\newpage
\section{Conclusion}
This paper makes the claim that the Decrits Consensus Algorithm is superior to POW algorithms in many ways: faster transaction confirmations, truly irreversible transactions, and permanent punishments to malevolent actors--all at virtually no external capital cost. At this time, the benefits of the DCA over other POS algorithms is not as clear. The Ripple consensus algorithm has a detailed paper describing its function, but the Ripple network operates on disparate principles that are difficult to compare. The author of this paper is unaware of any papers describing Peercoin, Nxt, or any other POS algorithm in enough detail to make a proper comparison.

The author invites the reader to comment, criticize, or question the contents of this paper at the official Decrits forum, \url{www.decrits.net}.

\newpage
\newgeometry{top=2cm}
\texttt{\small{
\hypertarget{pgp}
\\-----BEGIN PGP PUBLIC KEY BLOCK-----\linebreak
Version: GnuPG v2.0.22 (MingW32)\linebreak
\linebreak
mQGNBFRYf4kBDAD4dlaSDhZ31hAaoHBzOC2XvSJ9YVnsSA+s6q3quzdxfWLZF6Uu\\
wZZOItuSgH2Xzb3eDa0KwzaNOFCl8nSyyTeMUGqRYqIqo3KavDZj/4wELQ0JfF3n\\
Imcq4p8CsuixZXl8fW/XRYbAolHsy3/Z2tFLjFJjkMF8cB0MF5C0aNUSW7sQUIsd\\
ylIxF7IXuSwVCDJZQOBNmE9zOETkYpLZQZP2CvQqHdBdOsPxfubErP9WQhU4UPcd\\
vgR/LutNAlElw/UWL6J6ceb0rJSW5hDUArm5pqYymQpjqiCIq+jN8NxLLtjWuAcH\\
YSvKJgCL7c30Si8I84ghoPEb/462jmVcNd/htnqDmt7o+C38H6hcYcPCvWQBB0wB\\
xST3FF8p1SoqteZ5R4Okqz/DwFmXQkl4INHZw6mTISRxlgel2ZTN7snaxUfJTH15\\
xS9GrcPvTbmjnx8O4TWNMk3jveNXyM/tQSUrCsVd7DkwigWDa7rGTFT93+oiTb2l\\
1gV2vio6nlfCeEMAEQEAAbQiUnlhbiBQZmVpZmZlciA8cnlhbjI2aXhAeWFob28u\\
Y29tPokBuQQTAQIAIwUCVFh/iQIbAwcLCQgHAwIBBhUIAgkKCwQWAgMBAh4BAheA\\
AAoJEAmZHaRJVku/AW0MAPAr0F1pi7F7qU8qPIF9d4NGPbG4ysB6G77IOxJQBosJ\\
nmCCrtAObtusTrB1jsyP+RBryHHSCtnVlZGIlWG1p9+yQ9yLsONpPOZxRNtinH4a\\
x5Q0gO3H+JyuNYRVp9ugKerdLrRWY9G73WM3TEt2irle1bAupZQJw1LKHepa/xBc\\
03+bdgvZukmZQ2W530QzUgkWEfMXsjpZjW5RL6Ql/3h+IHqc/W64oaWI1RnbHc5c\\
dT/F4augtNOvXzSE/hb6YDNfwx3WcB0wxaKIhvhx6BJEU/2988MgLNGrpCltKpJs\\
JaiGYQ3+trsKJWosk5B8q5LBbOC3l1O4YlLCZBfjwWIPgE37DP2o+4j4plnEcic3\\
ZIPWIhqn62iaYzSL1uCnvaeDabAQzSVBbIKBnJNxZx1LKf5RrXmARNaO0H2qEQCR\\
7UhCcG4ldAY0o226421S94SKD52fZSNlYTxwpSDFLddQs90OtqJAxuauXAAu+gnP\\
Tjpi6XkxDMIRaITNW4PbkrkBjQRUWH+JAQwAzwJH4rQtUwNncO6oi3PJuIcPGH5b\\
iiANF8BLdtOP8fHn5I43eOwwtK82MQR2ZB8O4BzkABg7p2IAswuBOZxniI0ZftLi\\
TJ/LULUkGR+gwvEzJK/S1jY/j7+NatPKuOpp/3SL2D+cN5D0Gjni6u7nwdV7J0w3\\
tVrotoCG9hs097dmf70uvaDYoUTcUZbkf+Fiugv3YKdPnZ2Oqa8Nwf/XpBy1udct\\
Qq2GzfF4FLadUSsiM/QNlyZ6hlBM+6/SCUk49L3KSoyuHYMG5Cwu11xmeliIYv/K\\
HS/RhBYdDVtji5s/XHyjK6VSUCJ0hNQ/zgsCVtUisUXt4nLvNwvWi3ji1+MbnXBe\\
ePmSKw2KwT23MEsJHEOo1zd6A4CU1m1U6Zb8VkkAcM/0I7V/LfHUdaBKX5jAc+PV\\
d+yVHO2Sg7rEomSQBr6twPAl36vO1SKzH6yBvjlhXAZ3jFhaf16ECFnBHuODzX7O\\
JnZdTD0HUIODrUjdEWe5Ajb13kThcySasKNRABEBAAGJAZ8EGAECAAkFAlRYf4kC\\
GwwACgkQCZkdpElWS7/kDAv8CQZDECuYPr4ApdmNV2Ru/6y7EDYP0eBt4g/4KbcY\\
3WhikAL4ggdrxf7qDxmdWboLKKfgpT3VJjGBVliIBdHJUG5vd4oeNOpGBrpBXXaI\\
OhZSqRGJCeZS8Evk2cLGDNd0p4pLI5rs1c5QUhSrYqJsgl9MKnnFAIm8G/3elSw2\\
QlpiIR/HIP9xPu8T+qzlHbvBZUSya9oeL4A7JhoJ6wu7E949f7OlnlC17F0H5ZjL\\
XQbbb/3qCKasiMY3Ov0B/KIcbnphELKbo/hIrwdujFk+r8q+XjAyNOdaMtsnHjer\\
nGiUyjMRGqxG0Jmxlcxl4wr66b7l5fD84+38tuMkAZN5OnjsTHsTrZhlOaWbgYXZ\\
C/0lbxVsi28uAlyf6reqpoCguR6UwveGbS2fqu99wOigENb4Ik7tpp5QuckmZ3mG\\
VZUetangC7QJ5Df/iu6PNPAgE6tcwjIws5lzKIgQf4o8AitYZS7/+fj7cnt1v10i\\
Oo2W0FP5jn33vO9EBOvxD/Kk\\
=H0v+\\
-----END PGP PUBLIC KEY BLOCK-----}}
\end{document}